# TARGET SPEAKER EXTRACTION BY DIRECTLY EXPLOITING CONTEXTUAL INFORMATION IN THE TIME-FREQUENCY DOMAIN

*Xue Yang, Changchun Bao*, Jing Zhou, Xianhong Chen*

Institute of Speech and Audio Information Processing, Faculty of Information Technology,
Beijing University of Technology, Beijing 100124, China
yangx11@emails.bjut.edu.cn, baochch@bjut.edu.cn,
zhoujing@emails.bjut.edu.cn, chenxianhong@bjut.edu.cn

## ABSTRACT

In target speaker extraction, many studies rely on the speaker embedding which is obtained from an enrollment of the target speaker and employed as the guidance. However, solely using speaker embedding may not fully utilize the contextual information contained in the enrollment. In this paper, we directly exploit this contextual information in the time-frequency (T-F) domain. Specifically, the T-F representations of the enrollment and the mixed signal are interacted to compute the weighting matrices through an attention mechanism. These weighting matrices reflect the similarity among different frames of the T-F representations and are further employed to obtain the consistent T-F representations of the enrollment. These consistent representations are served as the guidance, allowing for better exploitation of the contextual information. Furthermore, the proposed method achieves the state-of-the-art performance on the benchmark dataset and shows its effectiveness in the complex scenarios.

***Index Terms***— Speech separation, target speaker extraction, speaker embedding, contextual information, attention mechanism

## 1. INTRODUCTION

In the real-world scenarios, the speech signals coming from different talkers often overlap, accompanied by background noise and reverberation. The human beings have the remarkable ability to focus on a target speaker, whereas the machines face a significant challenge in acquiring this capability. This well-known challenge is referred to as the cocktail party problem [1]. One potential approach to address this issue is through speech separation (SS), which aims to disentangle and separate all speech signals presented in the mixed signal [2]. Recently, the deep learning-based SS methods have demonstrated exceptional performance [3-4]. However, a practical difficulty with these methods is that they typically require prior knowledge of the number of speakers in the mixed signal. As an alternative solution to the cocktail party problem, the target speaker extraction (TSE) proves to be beneficial when the goal is to isolate the target speaker's speech exclusively. In order to achieve accurate extraction, it is crucial to provide auxiliary information that is specific to the target speaker. In this paper, the target speaker's enrollment is used as the auxiliary information due to its easy accessibility and effectiveness.

Numerous studies have been extensively explored to utilize target speaker's enrollment, where a majority of them rely on the speaker embedding. In many TSE methods, the enrollment is processed with a speaker embedder to obtain the fixed-dimensional speaker embedding for representing the characteristics of the target speaker. This speaker embedding is used to guide an extraction network for isolating the target speaker's speech from the mixed signal. In general, the TSE can be carried out either in the time-frequency (T-F) domain or in the time domain. The T-F domain approach involves the transformation of the mixed signal using short-time Fourier transform (STFT). In [5], the magnitude spectrum of the mixed signal was used to derive high-level features. These features are concatenated with the embedding obtained from a pre-trained speaker embedder for further extraction. In [6-8], the embedding derived from a jointly trained embedder was employed to avoid the potential sub-optimization arising from a pre-trained embedder on speaker verification or speaker recognition tasks. Besides, various techniques were adopted to enhance extraction performance, including the scaled activations [9], the speaker representation loss [10], the filter-like mask [11], etc. Nevertheless, the T-F domain approach typically utilizes the phase of the mixed signal to recover the target speaker's waveform, resulting in unsatisfactory outcomes. As a result, the time-domain approach becomes the mainstream since it directly processes the waveform of the mixed signal. In [12], the multi-scale features were extracted to capture the temporal structures at different resolutions. In [13], the weight sharing technique was applied to transform the mixed signal and the enrollment into the same feature space. Further improvements were achieved through the multi-stage structure [14], the iterative refined adaptation [15], the voice activity detection [16] and the advanced network architectures [17-19].

As a compact vector, the speaker embedding has been successfully employed in TSE. But it may not fully leverage the contextual information contained in the enrollment. Typically, the speaker embedding only contains speaker characteristics and omits the content details. However, these content details including local dynamics and temporal structures may be beneficial for TSE [20]. Consequently, there is a need to better exploit the contextual information contained in the enrollment for extending the speaker characteristics and including the content details. Recently, several studies have explored alternative methods to better utilize the enrollment. In [20-21], the feature sequences extracted from the enrollment and the mixed signal were combined and fused through various attention mechanisms. In [22], the target speaker's information was captured through the hidden state and cell state of the bi-directional long short-term memory (BLSTM) [23]. However, we argue that the target speaker's information may be partially lost as the extracted feature sequences or states are utilized.

Inspired by these embedding-free TSE methods [20-21] and the advanced SS techniques [24], we propose to directly exploit the contextual information within the enrollment in the T-F domain. Specifically, a simple and effective attention mechanism is employed to interact the T-F representations of the enrollment and the mixed signal for obtaining two weighting matrices. In these two weighting matrices, one is used for the real part and another one is used for the imaginary part of the complex spectrum. These two weighting matrices reflect the similarity among different frames of the T-F representations and are employed to obtain the weighted T-

F representations of the enrollment. These weighted representations, also referred as the consistent representations, have a spectrum pattern that is consistent with the mixed signal and are served as the guidance for further extraction. As a result, the direct interaction is achieved and the contextual information contained in the enrollment is better exploited. To the best of our knowledge, this is the first work that utilizes the direct interaction of the enrollment and the mixed signal in the T-F domain for the TSE. Furthermore, the experimental results demonstrate that the proposed method achieves the state-of-the-art (SOTA) performance on the benchmark dataset and exhibits the effectiveness in the complex scenarios.

The remainder of this paper is organized as follows. First, the proposed method is detailed in Section 2. Next, the experimental setup is presented in Section 3, followed by the analysis and discussions of the experimental results in Section 4. Finally, the conclusions are drawn in Section 5.

## 2. PROPOSED METHOD

### 2.1. Problem formulation

Given the target speaker's enrollment as the auxiliary information, the TSE aims at isolating the target speaker's speech from the mixed signal, namely:

$$\hat{x}_t = \mathcal{M}(y \mid e) = \mathcal{M}\left(\left(z_t + \sum_{k=1}^{S-1} z_k + n\right) \mid e\right) \quad (1)$$

where $\hat{x}_t$ is the estimated signal of the target speaker, $\mathcal{M}(A|B)$ denotes the mapping function of $A$ given $B$, $y$ is the mixed signal, $e$ represents the target speaker's enrollment, $z_t$ is the received signal of the target speaker, $z_k$ is the received signal of the $k^{\text{th}}$ interferer, $S$ represents the number of the speakers in the mixed signal and $n$ is the additive noise signal. In this paper, the estimated signal $\hat{x}_t$ corresponds to the clean speech of the target speaker for both the anechoic and reverberant scenarios.

### 2.2. Proposed network architecture

In this paper, the contextual information contained in the target speaker's enrollment is mainly exploited. Therefore, the proposed network is referred to as the CIENet for brevity. As illustrated in Fig. 1, the general framework of the CIENet consists of three modules, i.e., the encoder, the extractor and the decoder.

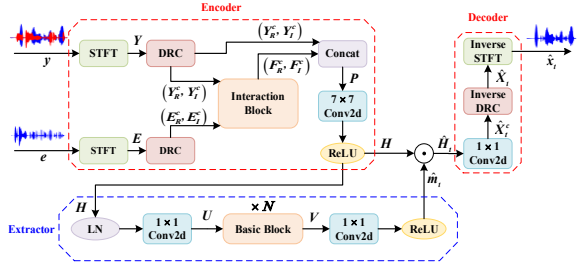

**Fig. 1**. The general framework of the proposed CIENet.

#### 2.2.1. The encoder

As depicted in the dotted box of the upper-left part of Fig. 1, the direct interaction between the enrollment and the mixed signal is achieved in the encoder. Both the enrollment $e$ and the mixed signal $y$ are served as the input of the encoder. These two waveforms are first transformed into their T-F representations $E$ and $Y$ through STFT, respectively. Note that the STFT is implemented with the convolutional layer for end-to-end (E2E) training. Next, the dynamic range compression (DRC) [25] is applied on the magnitude spectrum for leading to the compressed T-F representations $Y^c$ and $E^c$ expressed as follows:

$$Y^c = Y_R^c + jY_I^c = |Y|^\alpha \cos\theta_Y + j|Y|^\alpha \sin\theta_Y \quad (2)$$

$$E^c = E_R^c + jE_I^c = |E|^\alpha \cos\theta_E + j|E|^\alpha \sin\theta_E \quad (3)$$

where the features $Y_R^c \in R^{T_Y \times F}$ and $Y_I^c \in R^{T_Y \times F}$ are the real and imaginary parts of $Y^c$, respectively. Note that $T_Y$ is the frame number of the mixed signal and $F$ is the number of the frequency bins. The magnitude and the phase of the mixed signal are denoted as $|Y|$ and $\theta_Y$, respectively. Similarly, the features $E_R^c \in R^{T_E \times F}$ and $E_I^c \in R^{T_E \times F}$ are the real and imaginary parts of $E^c$. The frame number of the enrollment is denoted as $T_E$. The magnitude and the phase of the enrollment are $|E|$ and $\theta_E$. In addition, $j$ represents the imaginary unit. The compression factor $\alpha$ varies in the interval (0, 1].

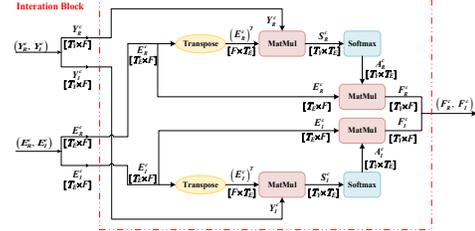

**Fig. 2**. The details of the interaction block in the encoder.

The interaction block takes the features $Y_R^c$, $Y_I^c$, $E_R^c$ and $E_I^c$ as its input and generates the weighted features $F_R^c$ and $F_I^c$ as the output. As illustrated in Fig. 2, the real-part features $Y_R^c$ and $E_R^c$ are interacted through a simple and effective attention mechanism. Specifically, the feature $E_R^c \in R^{T_E \times F}$ of the enrollment is transposed and matrix-multiplied with the feature $Y_R^c \in R^{T_Y \times F}$ of the mixed signal. This results in the matrix $S_R^c \in R^{T_Y \times T_E}$, which represents the similarity among different frames of these real-part features. To obtain the weighting matrix $A_R^c$, the softmax function is then applied to the last dimension of $S_R^c$. This weighting matrix $A_R^c$ is multiplied with the real-part feature $E_R^c$ to derive the weighted feature $F_R^c$ of the enrollment. A similar process is performed on the imaginary-part features to compute another feature $F_I^c$. This procedure can be formulated as follows:

$$F_R^c = Softmax\left(Y_R^c \left(E_R^c\right)^T\right) E_R^c \quad (4)$$

$$F_I^c = Softmax\left(Y_I^c \left(E_I^c\right)^T\right) E_I^c \quad (5)$$

where the superscript "$T$" denotes the transpose operation and $Softmax(\cdot)$ is the softmax function applied on the last dimension. For clarity, the shapes of various features are given in Fig. 2.

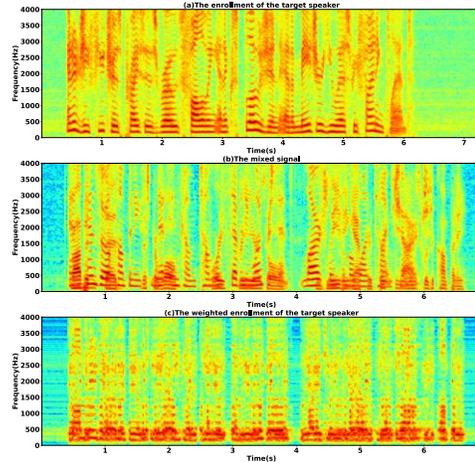

**Fig. 3**. The spectrums of (a) the enrollment of the target speaker; (b) the mixed signal; (c) the weighted enrollment of the target speaker.

To offer a clear view of the output from the interaction block, we present the spectrum of the weighted enrollment derived from the weighted features in Fig. 3(c). When compared with the spectrum of the mixed signal in Fig. 3(b), it becomes evident that the spectrum of the weighted enrollment not only retains the same number of frames as the mixed signal but also maintains a consistent spectrum pattern. Thus, the weighted T-F representations $F_R^c$ and $F_I^c$ are named as the consistent representations and can be employed as the guidance for better extraction. As indicated in Fig. 1, the features $Y_R^c$ and $Y_I^c$ of the mixed signal and the consistent representations of the enrollment are concatenated for forming a stacked feature tensor $P \in R^{4 \times T_Y \times F}$. This stacked tensor $P$ is processed by a convolutional layer and a rectified linear unit (ReLU) for the non-negative high-dimensional feature tensor $H \in R^{L \times T_Y \times F}$.

*2.2.2. The extractor and the decoder*

Within the extractor (dotted box at the bottom of Fig. 1), the non-negative feature tensor $H$ undergoes layer normalization (LN) and its channel dimension is changed from $L$ to $W$ through a convolutional layer. The resulting feature tensor $U \in R^{W \times T_Y \times F}$ is then processed through $N$ basic blocks. The output $V \in R^{W \times T_Y \times F}$ of these blocks is fed to another convolutional layer to restore the channel dimension to $L$, followed by a ReLU operation to estimate the mask $\hat{m}_t \in R^{L \times T_Y \times F}$ of the target speaker. This estimated mask is element-wise multiplied with the output $H$ of the encoder to obtain the feature tensor $\hat{H}_t \in R^{L \times T_Y \times F}$ of the target speaker.

Within the decoder (dotted box of the upper-right part of Fig. 1), this estimated feature tensor $\hat{H}_t$ is transformed back to the T-F representation $\hat{X}_t^c \in R^{2 \times T_Y \times F}$ through a convolutional layer. Besides, the inverse DRC and the inverse STFT are employed to reconstruct the target speaker's speech $\hat{x}_t$. Note that the inverse STFT is achieved through the transposed convolution for the E2E training.

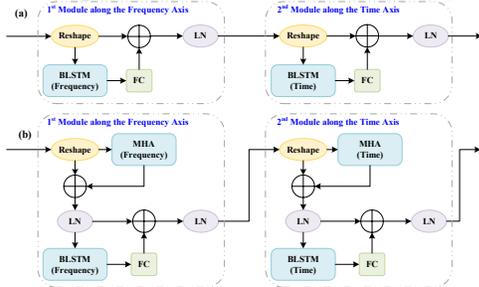

**Fig. 4**. (a) The mDPRNN block; (b) The mDPTNet block.

As illustrated in Fig. 4, two types of basic blocks are employed in the extractor. The first one is the modified dual-path recurrent neural network (mDPRNN), which is a T-F version of the block used in [3]. The mDPRNN block comprises two modules, one operates along the frequency axis and the other one operates along the time axis. Each module primarily consists of the BLSTM, fully connected (FC) layer, skip connection and LN. The second type of basic block is the modified dual-path transformer (mDPTNet), which is derived from the basic block in [4]. The mDPTNet block employs the multi-head attention (MHA) to better capture the global dependencies.

**2.3. Comparison with related work**

The most related work is the one that utilizes context-dependent bias [20]. However, our proposed CIENet differs from this work in the following ways: 1) In [20], the mixed signal and the enrollment are processed separately by a processing block to obtain the extracted feature sequences that are subsequently utilized to compute the context-dependent bias through an attention mechanism. Thus, there is a risk of partial loss of the target speaker's information. In contrast, our proposed CIENet achieves the direct interaction between the mixed signal and the enrollment in the T-F domain; 2) Only the magnitude spectrum is utilized in [20], whereas both the real and imaginary parts of the complex spectrum are employed in our proposed CIENet; 3) The auxiliary information of the non-target speaker is exploited in [20], while our focus is solely on the target speaker's enrollment.

Another related work is the speaker embedding-free target speaker extraction network (SEF-Net) [21] and the differences are as follows: 1) The SEF-Net is a time-domain approach, whereas our proposed CIENet performs extraction in the T-F domain; 2) The SEF-Net processes the mixed signal and the enrollment separately with the advanced conformer and then fuses the extracted feature sequences via cross MHA. In contrast, the CIENet achieves the direct interaction through an attention mechanism and processes the stacked T-F representations together after the concatenation; 3) The feature sequences derived from the mixed signal and the enrollment are fused multiple times in SEF-Net for different inter modules and intra modules. However, in our proposed CIENet, only one direct interaction is performed within the interaction block; 4) In SEF-Net, the padding or truncation is needed to make the frame number $T_E$ of the enrollment equal to the frame number $T_Y$ of the mixed signal, whereas such an operation is not required in the CIENet.

## 3. EXPERIMENTAL SETUP

**3.1. Datasets**

To assess the effectiveness of our proposed CIENet, three datasets are utilized and all derived from the Wall Street Journal (WSJ0) corpus. The first one is the WSJ0-2Mix [26], where two utterances of different speakers are mixed. The second dataset, WSJ0 Hipster Ambien Mixtures (WHAM!) [27], introduces noise to the clean mixture from WSJ0-2Mix. The third dataset WHAMR! [28] further takes the reverberation into account. The target speaker's enrollment is randomly selected for making it different from the utterance utilized in the mixing. Note that each speaker in the utterances is considered as the target speaker in turn. Besides, all speech signals are down-sampled to 8 kHz to reduce the computational complexity.

**3.2. Model configurations**

In this paper, the Hanning window with the length of 32ms is employed and the hop size is set to 16ms. The number $F$ of frequency bins is 129 and the compression factor $\alpha$ is set to 0.5. Additionally, the hyper-parameters $L$ and $W$ are set to 256 and 64, respectively. The number of hidden units in each direction of BLSTM is 128 and the number of attention heads is 4. Finally, the number $N$ of the basic blocks is 6.

**3.3. Training details**

In this paper, all models are trained for 120 epochs with 4s long speech signals. The Adam [29] optimizer is employed with an initial learning rate of 0.0005. The learning rate is multiplied by 0.98 for every two epochs during the first 100 epochs and multiplied by 0.9 in the last 20 epochs. The gradient clipping is applied for limiting the maximum $L_2$-norm to 1. The training objective is to maximize the scale-invariant signal-to-distortion ratio (SI-SDR) [30] between the estimated and the ground-truth signals of the target speaker.

## 4. RESULTS AND DISCUSSIONS

The extraction performance is evaluated through two well-known metrics, i.e., the SI-SDR improvement (SI-SDRi) and the SDR improvement (SDRi) [31].

**4.1. Comparison with the baseline methods**

In this subsection, several baseline methods are designed by following a similar general framework to our proposed CIENet. The differences are as follows: In Baseline 1, the interaction block is not utilized and a processing block with two BLSTM layers (128 hidden units for each direction) and a FC layer is employed. In this processing block, the magnitude spectrums of the enrollment and the mixed signal are processed separately to obtain two feature sequences. These sequences are utilized to derive the context-dependent bias through an attention mechanism [20]. Subsequently, the real and imaginary parts of the mixed signal, the feature sequence derived from the magnitude spectrum of the mixed signal and the context-dependent bias are concatenated along the channel dimension for further extraction; In Baseline 2, the same procedure as Baseline 1 is followed, except that the real and imaginary parts are utilized instead of the magnitude spectrums and two biases are obtained; In Baseline 3, the T-F features of the enrollment and the mixed signal are directly stacked along the channel dimension. Our proposed method is denoted as CIENet-mDPRNN when the mDPRNN block is utilized. Note that the processing block is not employed in the Baseline 3 and our CIENet-mDPRNN.

**Table 1**. Comparison with baseline methods on WSJ0-2Mix dataset.

| Methods | Params. (×10$^6$) | SI-SDRi (dB) | SDRi (dB) |
|---|---|---|---|
| Baseline 1 | 3.4 | 19.5 | 19.8 |
| Baseline 2 | 3.4 | 18.8 | 19.1 |
| Baseline 3 | 2.7 | 20.2 | 20.4 |
| CIENet-mDPRNN | 2.7 | 20.7 | 21.0 |

As given in Table 1, Baseline 1 can attain 19.5dB and 19.8dB on SI-SDRi and SDRi by leveraging both the context-dependent bias and advanced network architecture. Since this bias originates from the extracted feature sequences of the magnitude spectrums, the target speaker's information may be partially lost. In comparison, Baseline 2 exhibits performance degradation. This indicates that separately processing of the real and imaginary parts might not be a suitable choice. It is worth emphasizing that the processing block is exclusively employed in Baseline 1 and Baseline 2. Compared with these two methods, Baseline 3 surpasses 20.0dB on both metrics although there exist significant differences in the spectrum patterns as shown in Fig. 3(a) and Fig. 3(b). That is to say, Baseline 3 effectively utilizes the real and imaginary parts of the enrollment, discriminating the target speaker from the interferer. Note that the repetition or truncation operation is applied to the enrollment for ensuring $T_Y$ be equal to $T_E$. Furthermore, our proposed method demonstrates another 0.5dB improvements, highlighting the benefits of the consistent spectrum pattern, as indicated by Fig. 3(b) and Fig. 3(c). Besides, the number of parameters utilized in Baseline 3 and our proposed method is less.

### 4.2. Performance comparison on WSJ0-2Mix dataset

In this subsection, our proposed models are compared with several SOTA methods on WSJ0-2Mix dataset. Our proposed method is denoted as CIENet-mDPTNet when the mDPTNet block is utilized.

The majority of TSE methods (from SpeakerBeam + DC [9] to X-SepFormer ($S_{sc}$) [18] in Table 2) utilize an embedder to extract the target speaker's embedding. Among these embedding-based methods, SpeakerBeam + DC operating in the T-F domain attains 10.9dB on SDRi. This suboptimal result partially arises from its focus solely on the magnitude spectrums. In contrast, alternative methods demonstrate much higher performance in the time domain by directly processing the waveforms and implicitly considering the phase information. In SpEx [12] and SpEx+ [13], the convolutional layers are mainly employed and the receptive field is constrained. This leads to the scores lower than 18.0dB. In DPRNN-Spe-IRA [15], slightly better performance is achieved since the recurrent neural network is utilized and the speaker embedding is refined iteratively. In addition, SpEx++ [14], SpEx$_{pc}$ [17] and X-SepFomer ($S_{sc}$) [18] employ either the multi-stage architecture or advanced transformer for further improvement. Nevertheless, these speaker embedding-based methods may not fully exploit the contextual information contained in the enrollment. In comparison, the SEF-Net and VEVEN [22] exhibit the competitive scores without using speaker embedding. However, there is no direct interaction between the enrollment and the mixed signal in these two methods. In contrast, our proposed method achieves the direct interaction between the T-F representations through an attention mechanism. Furthermore, the proposed model, CIENet-mDPTNet, demonstrates 21.4dB and 21.6dB on SI-SDRi and SDRi, respectively.

**Table 2**. Comparison with SOTA methods on WSJ0-2Mix dataset.

| Methods | Type | Params. (×10$^6$) | SI-SDRi (dB) | SDRi (dB) |
|---|---|---|---|---|
| SpeakerBeam + DC [9] | T-F | — | — | 10.9 |
| SpEx [12] | Time | 10.8 | 16.6$^\dagger$ | 17.0$^\dagger$ |
| SpEx+ [13] | Time | 11.1 | 17.4 | 17.6 |
| DPRNN-Spe-IRA [15] | Time | 2.9 | 17.7 | 18.0 |
| SpEx++ [14] | Time | — | 18.0 | 18.4 |
| SpEx$_{pc}$ [17] | Time | 28.4 | 19.0 | 19.2 |
| X-SepFormer ($S_{sc}$) [18] | Time | — | 19.1 | 19.7 |
| SEF-Net [21] | Time | 27 | 17.2 | 17.6 |
| VEVEN [22] | Time | 2.6 | 19.0 | 19.2 |
| CIENet-mDPRNN | T-F | 2.7 | 20.7 | 21.0 |
| CIENet-mDPTNet | T-F | 2.9 | **21.4** | **21.6** |

-Results with superscript "$\dagger$" are given in [17].

### 4.3. Performance comparison on more complex datasets

As shown in Table 3, the proposed method is compared with SOTA methods on WHAM! and WHAMR! datasets. All four reference methods conducting TSE in the time domain showcase significant performance degradation in the presence of reverberation. In contrast, our proposed models not only achieve higher performance but also exhibit the effectiveness and robustness in reverberant scenarios. This stems from our method leveraging structured T-F representations and achieving improved guidance by thoroughly exploiting contextual information contained in the enrollment.

**Table 3**. Comparison with SOTA methods on complex datasets.

| Methods | WHAM! SI-SDRi (dB) | WHAM! SDRi (dB) | WHAMR! SI-SDRi (dB) | WHAMR! SDRi (dB) |
|---|---|---|---|---|
| SpEx [12] | 12.2$^\dagger$ | 13.0$^\dagger$ | 10.3$^\dagger$ | 9.5$^\dagger$ |
| SpEx+ [13] | 13.1$^\dagger$ | 13.6$^\dagger$ | 10.9$^\dagger$ | 10.0$^\dagger$ |
| DPRNN-Spe-IRA [15] | 14.2 | 14.6 | — | — |
| SpEx++ [14] | 14.3 | 14.7 | 11.7 | 10.7 |
| CIENet-mDPRNN | 15.7 | 16.1 | 15.5 | 14.1 |
| CIENet-mDPTNet | **16.6** | **17.0** | **15.7** | **14.3** |

-Results with superscript "$\dagger$" are given in [14].

### 5. CONCLUSIONS

In this paper, an improved TSE method in the T-F domain was proposed to fully exploit the contextual information contained in the enrollment. Specifically, an attention mechanism was employed to achieve the direct interaction between the T-F representations of the enrollment and the mixed signal. Two weighting matrices reflecting the similarity among different frames of the T-F representations were obtained and employed to yield the consistent representations of the enrollment. The resulting consistent representations exhibit a spectrum pattern consistent with that of the mixed signal and are served as the guidance for further extraction. In addition, the experimental results proved that the contextual information contained in the enrollment was better exploited through this direct interaction. With the mDPTNet block employed, our proposed CIENet-mDPTNet achieved the SOTA performance on WSJ0-2Mix dataset. Furthermore, our proposed method demonstrated its effectiveness and robustness in more complex scenarios.

### 6. ACKNOWLEDGEMENTS

This work was supported by the National Natural Science Foundation of China (Grants 61831019 and 62006010), R&D Program of Beijing Municipal Education Commission (KM202210005029).